\documentclass[12pt]{article}

\usepackage{graphicx,amsmath,amssymb,url,enumerate,mathrsfs,epsfig,color}
\usepackage{tikz}

\usepackage{pgfplots}

\usetikzlibrary{calc, patterns,arrows, shapes.geometric}
\pgfplotsset{compat=1.8}

\def\QED{\hskip0.1em\hfill\null\ \null\nobreak\hfill
\kern3pt\lower1.8pt\vbox{\hrule\hbox
{\vrule\kern1pt\vbox{\kern1.7pt \hbox{$\scriptstyle
QED$}\kern0.2pt}\kern1pt\vrule}\hrule}}
\newtheorem{defn}{Definition}[section]
\title{Hilbert bodies as quantum-classical continua}
\author{ Marcelo Epstein\footnote{University of Calgary, Canada.}}

\date{}
\begin{document}
\maketitle

\section{Motivation}

\begin{figure}[h]
\begin{center}

\begin{tikzpicture}

\draw[fill=white] (0,0)-- (3,0)--(4.5,2)--(1.5,2)--cycle;
\draw[-stealth'] (-2,2)--(-2,3);
\node[above] at (-2,3) {$z$};
\draw[-stealth'] (-2,2)--(-1,2);
\node[right] at (-1,2) {$y$};
\draw[-stealth'] (-2,2)--(-2.6,1.2);
\node[below left] at (-2.6,1.2) {$x$};

\draw[fill=white]  (7.5,2) to [out=120, in=180] (8,3.5) to [out=0, in=60] (8.5,2);

\draw[fill=white] (7.6,3.33)--(6.1,1.33)--(7,0)--(8.5,2);
\draw[fill=white]  (6,0) to [out=120, in=180] (6.5,1.5) to [out=0, in=60] (7,0);
\draw[ultra thick] (6,0)--(6.95,1.266);
\draw[ultra thick] (7,0)--(8.5,2);
\draw[ultra thick]  (3,0)--(4.5,2);
\draw[ultra thick] (0,0)--(1.5,2);
\end{tikzpicture}
\end{center}
\caption{Curvature-driven coupling}
\label{fig:example}
\end{figure}
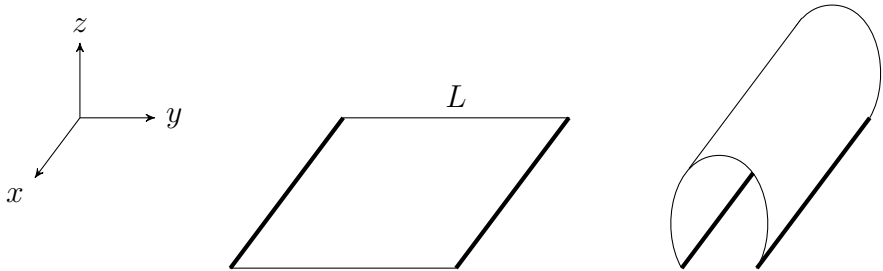

Graphene-based qubits constitute one of the most promising avenues for the realization of quantum computers. By rolling a graphene sheet to produce a nanotube, it is possible to enhance both the mechanical stability of the sheet and the temporal coherence of the qubits. As a thought experiment, imagine a  rectangular sheet ireinforced at two of its opposite edges with rigid bars, as shown in Figure \ref{fig:example}. As the two bars are moved in parallel towards each other, the sheet attains a curved shape, as shown. Suppose that a uniform distribution of qubits (say $\mu$ qubits per unit area) has been initially prepared uniformly over the whole sheet. If the sheet were to remain undeformed, the evolution would also be uniform and would abide by the Hamiltonian $H_0=\mu {\hat \sigma}_x$, where ${\hat \sigma}_x$ is the Pauli matrix in the $x$-direction. Remarking that this example is not representative of an actual physical system or a technology to realize it, we will assume that the Hamiltonian is affected by the local curvature according to $H_R=\mu (1+L/R) {\hat \sigma}_x$, where $R$ is the local radius of curvature. This radius depends on position and, by varying the distance between the rigid edges, it can be made to depend on time as well, thus achieving a kind of modulation.

Examples of this kind, where the qubits can be replaced with harmonic oscillators or more involved quantum units, make us think of the theory of continuous media with internal microstructure. It goes back to the pioneering work of the Cosserat brothers \cite{cosserat}, who proposed to enrich the kinematic description of a deformable body $\mathcal B$ by considering, as part of its constitution, additional degrees of freedom supplied by one or more vector fields defined over  $\mathcal B$. A granular medium, for example, can be conceived as a matrix (or macromedium) endowed at each point with a triple of linearly independent vectors representing the grains (or micromedium). The triad is permitted to undergo strictly linear deformations, implying that the grains are small enough so that they sustain states of constant strain. In modern terminology, one has replaced the body $\mathcal B$ (a 3-dimensional differentiable manifold) with its principal {\it frame bundle} $F{\mathcal B}$. A {\it configuration} of $F{\mathcal B}$ is a principal frame-bundle morphism $K: F{\mathcal B} \to F{\mathbb E}^3$, where $F{\mathbb E}^3$ is the natural frame bundle of the ordinary euclidean space of classical mechanics. Accordingly, we obtain the commutative diagram
\begin{equation}
\setlength{\unitlength}{2pt}
\begin{picture}(50,40)(0,0)
 \put(-2,28) {$F\mathcal B$}
  \put(2,23){\vector(0,-1){15}}
 \put(-7,15){$\pi_B$}
 \put(0,0){$\mathcal B$}
 \put(10,30){\vector(1,0){24}}
 \put(19,32){$K$}
 \put(37,28) {$F{\mathbb E}^3$}
 \put(40,23){\vector(0,-1){15}}
 \put(43,15){$\pi_E$}
 \put(38,0){${\mathbb E}^3$}
 \put(10,2){\vector(1,0){24}}
 \put(19.5,3){$\kappa$}
\end{picture}
\end{equation}
where $\kappa$ is an ordinary configuration of the macromedium, that is, an embedding of $\mathcal B$ into ${\mathbb E}^3$, and where $\pi_B$ and $\pi_E$ are the respective bundle projections of $F{\mathcal B}$ and $F{\mathbb E}^3$. We notice that, since bundle morphisms are, by definition, {\it fibre preserving}, the map $\kappa$ is already implied by the morphism $K$.

Other kinds of continua can be successfully modelled in this spirit \cite{capriz}. Thus, some theories of nematic liquid crystals use a single vector field to represent the internal structure, while A-smectics can be represented by means of a differential form, whose lack of local exactness indicates the presence of defects \cite{epsdual}. What is common to these models is that the corresponding microstructured bodies are represented by {\it associated bundles}\footnote{A rigorous treatment of the concept of associated bundles can be found, for example, in \cite{kobayashi}, pp 54-57.}  of $F{\mathcal B}$. A more general microstructure, whereby the body is an arbitrary fibre bundle not necessarily associated with the principal frame bundle of an ordinary body manifold, has been introduced in \cite{epsbuca, bucaeps}, on a purely theoretical basis.

In this paper, a model is proposed that can possibly accommodate a coupling between a classical continuum $\mathcal B$ and a quantum microstructure. This hybrid model is a fibre bundle over $\mathcal B$ whose typical fibre is a (complex) Hilbert space. As already remarked above, a technological motivation for such hybrids can be gathered from the many applications of graphene sheets, with a thickness consisting of a single atomic layer. Beyond such applications, however, is the idea of producing a viable geometric setting for a more general quantum-classical continuum. Although the Cosserat continuum has been invoked in a number of articles that include quantum-mechanical effects \cite{sikon1, sikon2, chervova, burnett}, the essential geometric structure consisting of a fibre bundle with a Hilbert-space fibre appears to be new and promising. It is perhaps worthwhile at the outset to state that, if any, this paper constitutes a contribution to continuum mechanics rather than to quantum physics. Interesting quantum-classical hybrids have been proposed (see, e.g. \cite{elze}) for discrete systems. Our objective here is only to provide a framework for incorporating some elements of the quantum paradigm into the classical continuum.\footnote{Hilbert bundles have been used to provide alternative formulations of quantum mechanics and quantum field theory. See \cite{prugovecki, drechsler, iliev1}.}

\section{Hilbert bodies}

Recall that a (smooth) fibre bundle consists of a differentiable manifold $\mathcal C$ (the {\it total space}), a {\it base manifold} $\mathcal B$, and a {\it typical fibre} manifold $\mathcal F$. These three manifolds are related in the following way:
\begin{enumerate}
\item There is a distinguished differentiable surjective {\it projection map} $\pi: {\mathcal C} \to {\mathcal B}$ of everywhere maximal rank. In other words, the projection map is a surjective submersion. At each point $b \in {\mathcal B}$, the set $\pi^{-1}(\{b\})$ is called the {\it fibre over $b$}.
\item Every point $b \in {\mathcal B}$ has an open neighbourhood $V \subset {\mathcal B}$ and a diffeomorphism $\phi_V:\pi^{-1}(V) \to V \times {\mathcal F}$ such that the following diagram is commutative:
\begin{equation}
\setlength{\unitlength}{2pt}
\begin{picture}(50,40)(0,0)
 \put(-10,28) {$\pi^{-1}(V)$}
 \put(10,30){\vector(1,0){24}}
 \put(17,32){$\phi_V$}
 \put(39,28) {$V \times \mathcal F$}
 \put(40,23){\vector(0,-1){15}}
 \put(43,15){$pr_1$}
 \put(39,0){$V$}
 \put(10,23){\vector(3,-2){25}}
 \put(13,13){$\pi$}
\end{picture}
\end{equation}
where $pr_1$ stands for the first projection of a Cartesian product. The pair $(V, \phi_V)$ is called a {\it local trivialization}.
\item Whenever two local trivializations, $(V_1, \phi_1)$ and $(V_2,\phi_2)$, have a non-empty intersection $V=V_1 \cap V_2 \ne \emptyset$, the {\it transition map} $\phi_{1,2}=\phi_2 \circ \phi_1^{-1}:\phi_1(V) \to \phi_2(V)$ is a diffeomorphism. We notice that at each point $b \in V$ the transition map is a diffeomorphism of the typical fibre onto itself. We require each of these fibre diffeomorphisms to belong to a Lie group of transformations $\mathcal G$, known as the {\it structure group} of the bundle. This group is part and parcel of the definition.
\end{enumerate}
More intuitively, a fibre bundle looks locally (that is, chunk-wise) as the product of two manifolds, namely, $V \times {\mathcal F}$. This identification is called local triviality. Once a trivialization is given, however, any other trivialization must be compatible with it, in the sense that the relation between the trivializations is restricted to a group of transformations $\mathcal G$ of the typical fibre $\mathcal F$. For example, if $\mathcal F$ is a vector space, we may require that $\mathcal G$ be the group of linear automorphisms of $\mathcal F$.

\begin{defn} {\rm A {\it Hilbert bundle} $\mathcal H$ is a fibre bundle over an ordinary body $\mathcal B$ whose typical fibre is a (separable) complex Hilbert space $H$ and whose structure group is the unitary group $\mathcal U_H$ of $H$.}
\end{defn}

We can construct the principal bundle ${\mathcal P}$ over $\mathcal B$ associated with $\mathcal H$ in the standard fashion \cite{schottenloher} by construing its fibre as the unitary group $\mathcal{ U}_H$. As in any principal bundle, this group has a {\it right} action on the bundle itself. We will refer to this principal bundle as a {\it Hilbert body}. The fibre at $X \in {\mathcal B}$ will be denoted by ${\mathcal P}_X$.

\section{Configurations and deformations}

Our Hilbert body manifests itself in the Cartesian product ${\mathcal S}={\mathbb E}^3 \times {\mathcal U}_ H$, in the sense suggested by the following definition.

\begin{defn} {\rm A {\it configuration} of a Hilbert body $\mathcal P$ is a fibre-preserving embedding $K: {\mathcal P} \to {\mathcal S}$.}
\end{defn}

Given two configurations, $K_0$ and $K$, the {\it deformation} from the first to the second is the composition
\begin{equation}
\Xi = K \circ K_0^{-1}:K_0({\mathcal P}) \to K({\mathcal P}).
\end{equation}
As a fibre-preserving map, $\Xi$ implies also an ordinary deformation $\xi$ of the base manifold $\mathcal B$. A deformation $\Xi$ can be clearly regarded as a fibre-bundle morphism. The fibre-wise maps are assumed to be unitary transformations. Thus, the time evolution of a Hilbert body consists of an ordinary classical mechanics deformation of the body $\mathcal B$ supplemented with a quantum field riding on the fibres. The interaction between these two mechanisms is the subject of the constitutive equations that define the particular nature of each system.

In a first-grade theory (for the so-called {\it simple bodies}) the material response depends only on the local 1-jet of the deformation. At each point $X \in {\mathcal B}$, this 1-jet consists of 4 elements: (i) the macroscopic deformation $\xi(X)$, which we discard given the standard assumed invariance under spatial translations; (ii) the classical deformation gradient, which is a linear isomorphism ${\bf F}(X):T_X{\mathcal B} \to T_{\xi(X)}{\mathbb E}^3$; (iii) a unitary transformation $U(X):{\mathcal P}_X \to {\mathcal P}_{\xi(X)}$; (iv) the referential gradient $\nabla U (X)$, which, in a coordinate system $X^I (I=1,2,3)$ of $\mathcal B$ in the reference configuration $K_0$, can be expressed as the 3 partial derivatives $U_{,I}$.

\section{Constitutive equations and time evolution}
The kinematic setting introduced above appears to be general enough to accommodate the most sophisticated theories, including quantum field theories and entanglement, but this is not the intention of of the present work. Rather, the objective is to provide a geometric framework to allow for a modicum of coupling between a classical macromedium and a quantum micromedium. 

With this idea in mind, we can assume that the fibre at each point abides by the Schr${\rm \ddot o}$dinger equation, except that the parameters of the system at hand depend on the present state of deformation of the macromedium at that point. Thus, the evolution is that of a system with time-dependent parameters,\footnote{See, for example \cite{villasenor} for a detailed treatment of the time-dependent quantum harmonic oscillator.}  although the time-dependence is not explicit, but mediated by the current value the deformation gradient ${\bf F}(X)$ via, for instance, the invariants of the tensor ${\bf C=F^TF}$.

Concomitantly, we may be at liberty of assuming that the deformation of the macromedium is also affected by the present state of the micromedium. This issue, however, may be problematic, since even an indirect measurment of the state of a quantum system leads to a collapse to an eigenvector. A milder, perhaps permissible, coupling could be obtained through the gradient $\nabla U$. For each value of the index $I$, the product $W_I=U^\dagger U_{,I}$ is an antihermitian operator, which can be used to mediate the coupling by means, for example, of the norm of the resulting trace vector ${\rm tr } W_I$.

Perhaps the simplest example of a Hilbert body consists of identifying the typical fibre $H$ with the configuration space of a qubit, thus rendering a finite-dimensional fibre. A time-dependent coupling can be achieved by multiplying each of the Pauli matrices by a real function of the tensor $\bf C$, while, if necessary, the elasticity of the underlying classical body can be made to depend on the norm alluded to above. Applicability of such systems can be substantiated experimentally, as shown in \cite{clercq}. Vibrations of a graphene sheet can better be modelled with the infinite dimensional fibre of the harmonic oscillator.

\end{document}